# Predicting Competitive and Non-Competitive Torquoselectivity in Ring-Opening Reactions using QTAIM and the Stress Tensor

Alireza Azizi[1], Roya Momen[1], Alejandro Morales-Bayuelo[2,3], Tianlv Xu[1], Steven R. Kirk[1*] and Samantha Jenkins[1]*

[1]*Key Laboratory of Chemical Biology and Traditional Chinese Medicine Research and Key Laboratory of Resource Fine-Processing and Advanced Materials of Hunan Province of MOE, College of Chemistry and Chemical Engineering, Hunan Normal University, Changsha Hunan 410081, China.*

[2]*Grupo de Investigaciones Básicas y Clínicas de la Universidad del Sinú, escuela de medicina, Cartagena de Indias, Colombia*

[3]*Centro de Investigación de Procesos del Tecnológico Comfenalco. Programa de Ingeniería Industrial. Fundación Universitaria Tecnológico Comfenalco, Cartagena de Indias, Colombia.*

e-mail: steven.kirk@cantab.net
e-mail: samanthajsuman@gmail.com

We present a new vector-based representation of the chemical bond referred to as the bond-path frame-work set $\mathbb{B} = \{p, q, r\}$, where $p$, $q$ and $r$ represent three paths with corresponding eigenvector-following path lengths $\mathbb{H}^*$, $\mathbb{H}$ and the bond-path length from the quantum theory of atoms in molecules (QTAIM). We find that longer path lengths $\mathbb{H}$ of the ring-opening bonds predict the preference for the transition state inward (TSIC) or transition state outward (TSOC) ring opening reactions in agreement with experiment for all five reactions **R1-R5**. Competitiveness and non-competitiveness have traditionally been considered using activation energies. The activation energy however, for **R3** does not satisfactorily determine competitiveness or provide consistent agreement with experimental yields. We choose a selection of five competitive and non-competitive reactions; methyl-cyclobutene (**R1**), ethyl-methyl-cyclobutene (**R2**), iso-propyl-methyl-cyclobutene (**R3**), ter-butyl-methyl-cyclobutene (**R4**) and phenyl-methyl-cyclobutene (**R5**). Therefore, in this investigation we provide a new criterion, within the QTAIM framework, to determine whether the reactions **R1-R5** are competitive or non-competitive. We that find **R2**, **R3** and **R5** are competitive and **R1** and **R4** are non-competitive reactions in contrast to the results from the activation energies, calling into question the reliability of activation energies.

# 1. Introduction

Torquoselectivity is defined to be the preference for either the transition state (TS) inward conrotatory reaction pathway (TSIC) or the transition state outward conrotatory (TSOC) reaction pathway.[1,2] Houk[3] first postulated the mechanism of torquoselectivity using orbital symmetry to understand electrocyclic reactions where high symmetry is present.[4–11] In reactions where the electrostatic and steric effects are more important that the symmetry parameters however, the traditional formalism based on the Woodward-Hoffmann rules is insufficient.[12–14] In this investigation therefore, we select five ring-opening reactions where the Frontier Molecular Orbital (FMO) theory has not been satisfactory (see **R1**[15] and **R2-R5**[16]) in explaining the reaction mechanism; 3-methyl-cyclobut-1-ene (**R1**), 3-(ethyl,methyl)-cyclobut-1-ene (**R2**), 3-(isopropyl,methyl)-cyclobut-1-ene (**R3**), 3-(terbutyl,methyl)-cyclobut-1-ene (**R4**) and 3-(phenyl, methyl)-cyclobut-1-ene (**R5**), see **scheme 1** and **Table 1**. We choose a series of reactions **R1**-**R5** with contrasting nuclear skeletons and activation energies, conventionally considered to be either competitive or non-competitive on the basis of ΔTS < 1.0 kcal/mol, see **Scheme 1, Table 1** and **Figure 1**.

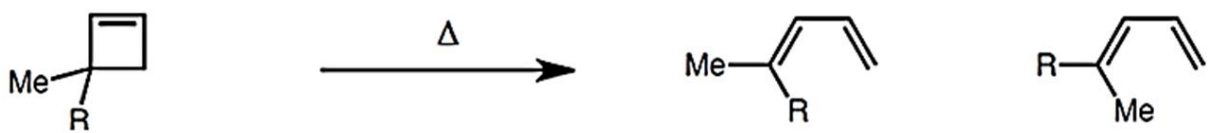

**Scheme 1.** Conrotatory ring-opening reactions with TSIC and TSOC reaction pathways with (**R1**); R = H, (**R2**); R = ethyl, (**R3**); R = isopropyl, (**R4**); R = terbutyl and (**R5**); R = phenyl, see the main text for further explanation.

Recent studies have shown new interpretations to treat this bottleneck in electrocyclic reactions using frameworks ranging from the Electron Localization Function (ELF),[17,18] electronic structure principles,[19] to stress tensor and QTAIM.[20–22] Of these formalisms, only the stress tensor and QTAIM[22] allowed the identification and quantification of the reaction coordinate through directional parameters, although the later was limited to use with reactions with very similar nuclear skeletons.

**Table 1**. The **ΔTS (TSIC-TSOC)** activation energies (kcal/mol) obtained using b3lyp/6-31G(d,p) theory level, see **Figure 2**, for the ring-opening reactions **R1**-**R5**[3,22,23]. The cut-off for stereospecificity is defined to be for experiment yields of approximately 70% for **R1**[15] and for **R2-R5**[16], the dominant product is displayed for clarity in a bold font.

| Reaction | Reactant | ΔTS (TSIC-TSOC) | Product and yield (%) | |
|---|---|---|---|---|
| | | | TSIC | TSOC |
| R1 | 3-methyl-cyclobut-1-ene[a] | +6.040 | Z-methyl-2,4-pentadiene (8%) | **E-methyl-2,4-pentadiene (83%)** |
| R2 | 3-(ethyl,methyl)-cyclobut-1-ene | -0.508 | **Z-(ethyl,methyl)-3,5-hexadiene (68%)** | E-(ethyl,methyl)-3,5-hexadiene (32%) |
| R3 | 3-(isopropyl,methyl)- cyclobut-1-ene | +1.308 | **Z-(isopropyl,methyl)-3,5-hexadiene (65%)** | E-(isopropyl,methyl)-3,5-hexadiene (35%) |
| R4 | 3-(terbutyl,methyl)-cyclobut-1-ene | +1.419 | Z-(terbutyl,methyl)-3,5-hexadiene (32%) | **E-(terbutyl,methyl)-3,5-hexadiene (68%)** |
| R5 | 3-(phenyl,methyl)-cyclobut-1-ene | +3.324 | Z-(phenyl,methyl)-3,5-hexadiene (30%) | **E-(phenyl,methyl)- 3,5-hexadiene (70%)** |

[a] Experiment yield for 90% of the completion of the reaction.

It can be seen from **Table 1** that there is a discrepancy for **R3** between the preference for TSOC from the calculated ΔTS(TSIC-TSOC) yield and the experimental yield, see **Figure 2(c)**. The larger yield (**65%**) for the **TSIC** experiment product indicated that it is preferred over the smaller yield of (35%) for the TSOC product. The recent investigation by some of the current authors successfully demonstrated a new method to distinguish the torquoselectivity along the reaction coordinate of a pair of competitive reactions without the use of orbital information or symmetry.[22] There are however, disadvantages of this recent QTAIM and stress tensor analysis. The first was that the analysis was restricted to sets of TSIC/TSOC competitive reactions with very similar nuclear skeletons. The second was that the analysis was complex and required to be performed on all bonds, i.e. bond critical points (*BCPs*) in each structure.

The first goal of this current investigation is consider non-competitive reactions in addition to competitive reactions. In particular, we seek to find a new criterion to determine whether a reaction is competitive or non-competitive based on chemical properties rather than the somewhat arbitrary ΔTS < 1.0 kcal/mol, see **Table 1**. The second goal and main goal of this investigation will be to discover a method to distinguish the torquoselectivity along the reaction coordinate of competitive reactions as well as non-competitive reactions that accommodates the directional character of the reaction pathways without the use of orbital information or symmetry. Then we can address the problem with the discrepancy seen for **R3** in **Table 1** for the TSIC and TSOC calculated ΔTS(TSIC-TSOC) barriers with the yield of the product.

## 2. Theory and Methods

*2.1 The QTAIM and stress tensor BCP descriptors; the metallicity $\xi(r_b)$, the stress tensor polarizability $\mathbb{P}_\sigma$*

As a topological interpretation tool for the extension of quantum mechanics, the Quantum Theory of Atoms in Molecules (QTAIM)[24] provides methods to study the electron distribution $\rho(\mathbf{r})$, returning a rich quantitative description of bonding environments of the atoms in a molecule. We can define a piecewise continuous gradient path in the scalar field $\rho(\mathbf{r})$ by evaluating $\nabla\rho(\mathbf{r})$ at some point, then following this vector for an extremely small distance and evaluating $\nabla\rho(\mathbf{r})$ again.

An atom can be defined as a region of real space bounded by surfaces through which there is zero flux in the gradient vector field of the total electronic charge density distribution. An interaction surface is defined by the set of trajectories that terminate at the point where $\rho(\mathbf{r}) = 0$. This means that an interatomic surface satisfies the "zero-flux" boundary conditions: $\nabla\rho(\mathbf{r})\cdot\mathbf{n}(\mathbf{r}) = 0$

Where $\mathbf{n}(\mathbf{r})$ is the unit vector common to the surface at $\mathbf{r}$, so the surface is not crossed by any of the trajectories of $\rho(\mathbf{r})$. The charge density $\rho(\mathbf{r})$ is a physical quantity which has a definite value at each point in space and whose form is controlled by the forces exerted on it by the nuclei, so its topological structure is comparatively simple. Analyzing the properties of the Hessian matrix of $\rho(\mathbf{r})$ at each critical point allow us to recognize different types. A diagonalization of the Hessian matrix of $\rho(\mathbf{r})$ gives the ordered eigenvalue set $\lambda_1 < \lambda_2 < \lambda_3$, with the Laplacian of the electron density being the algebraic sum of these eigenvalues. These eigenvalues are associated with a corresponding set of eigenvectors $\underline{\mathbf{e}}_1$, $\underline{\mathbf{e}}_2$, $\underline{\mathbf{e}}_3$. The analysis of the eigenvalues of the Hessian matrix of $\rho(\mathbf{r})$ indicate whether the forces exerted by and on the $\rho(\mathbf{r})$ favors tensile modes, i.e. expansion or compression for the volume element according to the positive or negative eigenvalue, respectively.

In the limit that the forces on the nuclei become vanishingly small, an atomic interaction line (AIL)[25] becomes a bond-path, although not necessarily a chemical bond.[26]

The curvature of charge density $\rho(\mathbf{r})$ can provide insights into the bonding situation. This curvature can expressed by an examination of the proportion of the eigenvalues, named as bond ellipticity. The $\lambda_1$ and $\lambda_2$ eigenvalues are the negative eigenvalues of the Hessian matrix and are perpendicular to the bond axis. The ellipticity is expressed as: $\varepsilon = |\lambda_1|/|\lambda_2| - 1$.[24] The presence of a degree of covalent character is determined from the total local energy density $H(\mathbf{r}_b)$,[21,27] which is defined as:

$$H(\mathbf{r}_b) = G(\mathbf{r}_b) + V(\mathbf{r}_b) \tag{1}$$

In equation **(1)**, $G(\mathbf{r}_b)$ and $V(\mathbf{r}_b)$ are the local kinetic and potential energy densities at a *BCP*, respectively. A value of $H(\mathbf{r}_b) < 0$ for the closed-shell interaction, $\nabla^2\rho(\mathbf{r}_b) > 0$, indicates a *BCP* with a degree of covalent character and conversely $H(\mathbf{r}_b) > 0$ reveals a lack of covalent character for the closed-shell *BCP*. A cross-section taken through the charge density, at a *BCP* with ellipticity $\varepsilon = 0$, perpendicular to the bond-path, would reveal a circular distribution of electronic charge density. A related quantity to the ellipticity $\varepsilon$ for closed-shell interactions is the metallicity, $\xi(\mathbf{r}_b) = \rho(\mathbf{r}_b)/\nabla^2\rho(\mathbf{r}_b) \geq 1$, where $\rho(\mathbf{r}_b)$ and $\nabla^2\rho(\mathbf{r}_b)$ are the values of total electronic charge density and the Laplacian respectively, at the *BCP*. The metallicity $\xi(\mathbf{r}_b)$,[28,29] previously has been used to explore suspected metallicity ranges of metals, metalloids and non-metals.[28,29] Some of the current authors also the metallicity $\xi(\mathbf{r}_b)$ measure by showing that the $\xi(\mathbf{r}_b)$ is inversely related to "nearsightedness" of the first-order density matrix and is suitable for closed-shell systems.[30] The metallicity $\xi(\mathbf{r}_b)$ has recently been shown to be important for the ring-opening *BCP*.[22] The metallicity $\xi(\mathbf{r}_b)$ will therefore relate to the reaction electronic flux (REF, $J(\xi)$) corresponding to a chemical process along the IRC($\xi$) is defined as $J(\xi) = -d\mu/d\xi$,[31] where $\mu$ is the chemical potential. The use of the same Greek letter 'ξ' for the definitions of metallicity $\xi(\mathbf{r}_b)$ and REF $J(\xi)$ is coincidental.

The stress tensor descriptors are based on forth order derivatives of the total charge density distribution $\rho(\mathbf{r}_b)$ since they are calculated from the Hessian matrix of the stress tensor.[32,33] Previously, the stress tensor polarizability $\mathbb{P}_\sigma = |\lambda_{3\sigma}|/|\lambda_{1\sigma}|$ was defined as the reciprocal of the stress tensor stiffness $\mathbb{S}_\sigma$.[34] Larger values of the stress tensor polarizability $\mathbb{P}_\sigma$ indicate the dominance of the tensile eigenvalue $\lambda_{3\sigma}$ compared $\lambda_{1\sigma}$ which is the compressive and therefore corresponds to greater bond-path $\mathbb{P}_\sigma = |\lambda_{3\sigma}|/|\lambda_{1\sigma}|$.

## 2.2 The QTAIM bond-path properties; BPL, the eigenvector-following path lengths $\mathbb{H}$, $\mathbb{H}^*$ and the bond-path framework set $\mathbb{B}$

The bond-path length (BPL) is defined as the length of the path traced out by the $\underline{\mathbf{e}}_3$ eigenvector of the Hessian of the total charge density $\rho(\mathbf{r})$, passing through the *BCP*, along which $\rho(\mathbf{r})$ is locally maximal with respect to any neighboring paths. The bond-path curvature separating two bonded nuclei is defined as the dimensionless ratio:

(BPL - GBL)/GBL, (2)

Where BPL is as the associated bond-path length and the geometric bond length (GBL) is the inter-nuclear separation. The BPL often exceeds the GBL particularly for weak or strained bonds and unusual bonding environments.[35] Earlier, one of the current authors hypothesized that a bond-path may possess 1-D, 2-D or a 3-D morphology[36,37] with 2-D or a 3-D bond-paths associated with a *BCP* with ellipticity $\varepsilon > 0$, being due to the differing degrees of charge density accumulation, of the $\lambda_2$ and $\lambda_1$ eigenvalues respectively. Bond-paths possessing zero and non-zero values of the bond-path curvature defined by equation **(2)** can be considered to possess 1-D and 2-D topologies respectively. For the realization of this hypothesis we start by choosing the length traced out in 3-D by the path swept by the tips of the scaled $\underline{\mathbf{e}}_2$ eigenvectors of the $\lambda_2$ eigenvalue, the scaling factor could be the ellipticity $\varepsilon$. For shared-shell *BCP*s the value of ellipticity $\varepsilon$ correlates with single bonds for approximately $\varepsilon \leq 0.1$ and double bonds for $\varepsilon \geq 0.25$, however for closed-shell *BCP*s there is no such correlation for the chemical character of a bond. With *n* scaled eigenvector $\underline{\mathbf{e}}_2$ tip path points $q_i = r_i + \varepsilon_i \underline{\mathbf{e}}_{2,i}$ on the path *q* where $\varepsilon_i$ = ellipticity at the $i^{th}$ bond-path point $r_i$ on the bond-path *r*. It should be noted that the bond-path is associated with the $\lambda_3$ eigenvalues of the $\underline{\mathbf{e}}_3$ eigenvector does not take into account differences in the $\lambda_1$ and $\lambda_2$ eigenvalues of the $\underline{\mathbf{e}}_1$ and $\underline{\mathbf{e}}_2$ eigenvectors. Analogously, for the $\underline{\mathbf{e}}_1$ tip path points we have $p_i = r_i + \varepsilon_i \underline{\mathbf{e}}_{1,i}$ on the path *p* where $\varepsilon_i$ = ellipticity at the $i^{th}$ bond-path point $r_i$ on the bond-path *r*.

We will refer to the new QTAIM interpretation of the chemical bond as the *bond-path framework set* that will be denoted by $\mathbb{B}$, where $\mathbb{B} = \{p, q, r\}$. This effectively means that in the most general case a bond is comprised of three 'linkages'; *p*, *q* and *r* associated with the $\underline{\mathbf{e}}_1$, $\underline{\mathbf{e}}_2$ and $\underline{\mathbf{e}}_3$ eigenvectors, respectively. The *p* and *q* parameters define eigenvector-following path length $\mathbb{H}$ and $\mathbb{H}^*$, see **Scheme 2**:

$$\mathbb{H}^* = \sum_{i=1}^{n-1} |p_{i+1} - p_i| \quad (3a)$$

$$\mathbb{H} = \sum_{i=1}^{n-1} |q_{i+1} - q_i| \quad (3b)$$

The *eigenvector-following path* length $\mathbb{H}^*$ or $\mathbb{H}$ refers to the fact that the tips of the scaled $\underline{\mathbf{e}}_1$ or $\underline{\mathbf{e}}_2$ eigenvectors will sweep out along the extent of the bond-path, defined by the $\underline{\mathbf{e}}_3$ eigenvector, between the two bonded nuclei that the bond-path connects. In the limit of vanishing ellipticity $\varepsilon = 0$, *for all* steps $i$ along the bond-path, one has $\mathbb{H}$ = BPL and $\mathbb{H}$ > BPL.

From the form of $p_i = r_i + \varepsilon_i \underline{\mathbf{e}}_{1,i}$ and $q_i = r_i + \varepsilon_i \underline{\mathbf{e}}_{2,i}$ we see for shared-shell *BCP*s, that in the limit of the ellipticity $\varepsilon \approx 0$ i.e. corresponding to single bonds, we then have $p_i = q_i = r_i$ and therefore the value of the lengths $\mathbb{H}^*$ and $\mathbb{H}$ attain their lowest limit; the bond-path length (*r*) BPL. Conversely, higher values of the ellipticity $\varepsilon$, for instance, corresponding to double bonds will always result in values of $\mathbb{H}^*$ and $\mathbb{H}$ > BPL. Additionally, because $\mathbb{H}$ and $\mathbb{H}^*$ are defined by the distances swept out by the $\underline{\mathbf{e}}_2$ tip path points, $q_i = r_i + \varepsilon_i \underline{\mathbf{e}}_{2,i}$ and $p_i = r_i + \varepsilon_i \underline{\mathbf{e}}_{1,i}$ respectively, one has $\mathbb{H} = \mathbb{H}^*$ provided that identical scaling factor $\varepsilon_i$ is used in equations **(3a)** and **(3b)**.

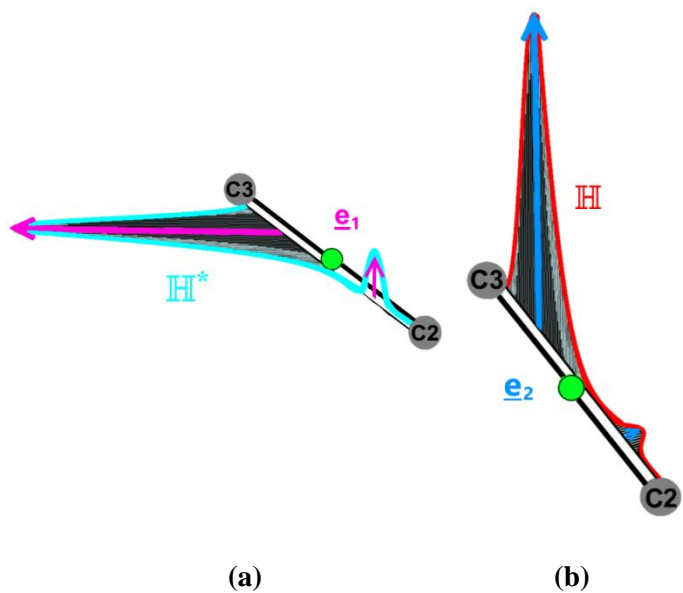

(a)    (b)

**Scheme 2.** The pale blue line in sub-figure **(a)** represents the path, referred to as the eigenvector-following path length $\mathbb{H}^*$, swept out by the tips of the scaled $\underline{\mathbf{e}}_1$ eigenvectors, shown in magenta, and defined by equation **(3a)**. The red path in sub-figure **(b)** corresponds to the eigenvector-following path length $\mathbb{H}$, constructed from the path swept out by the tips of the scaled $\underline{\mathbf{e}}_2$ eigenvectors, shown in mid blue and is defined by equation **(3b)**. The pale and mid blue arrows representing the $\underline{\mathbf{e}}_1$ and $\underline{\mathbf{e}}_2$ eigenvectors are scaled by the ellipticity $\varepsilon$ respectively, where the vertical scales are exaggerated for visualization purposes. The green sphere indicates the position of a given *BCP*. Details of how to implement the calculation of the eigenvector-following path lengths $\mathbb{H}^*$ and $\mathbb{H}$ are provided in the **Supplementary Materials S6**.

A bond within QTAIM is defined as being the bond-path traversed along the $\underline{\mathbf{e}}_3$ eigenvector of the $\lambda_3$ eigenvalue from the bond-path, but, as a consequence of equation **(3)**, this definition should be expanded. The new definition of a bond should consider the bond-path to comprise the two paths swept out by the $\underline{\mathbf{e}}_2$ and $\underline{\mathbf{e}}_1$ eigenvectors that form the eigenvector-following path length $\mathbb{H}$ and $\mathbb{H}^*$, respectively.

In this investigation we shall compare the eigenvector-following path lengths $\mathbb{H}$ of the shared-shell ring-opening *BCP*s of the bond-paths of the TSIC and TSOC reaction pathways for **R1**-**R5**, see **Figure 1** and **Figure 6.** As was mentioned previously, the ellipticity $\varepsilon_i$ values along the bond-paths ($r$) associated with shared-shell *BCP*s are chemically meaningful, i.e. indicating single or double bond character unlike the corresponding values for closed-shell *BCP*s. This is being undertaken to determine the preference for either the transition state (TS) inward conrotatory reaction pathway (TSIC) or the transition state outward conrotatory (TSOC). Therefore, we will examine the bond-paths of the shared-shell ring-opening *BCP*s of the TSIC and TSOC reaction pathways before they transform to closed-shell *BCP*s or rupture entirely to obtain predictions of the TSIC or TSOC product preference. As a consequence, we suggest longer $\mathbb{H}$ lengths associated with the shared-shell ring opening *BCP*s predict the preferred TSIC or TSOC reaction pathway. This is because longer $\mathbb{H}$ lengths for a given bond-path ($r$) are due to the presence of higher ellipticity $\varepsilon$ values, these in turn indicate stronger shared-shell ring-opening *BCP*s during the reaction and therefore predict the preferred TSIC or TSOC product.

## 3. Computational Details

Transition states were optimized and checked for the presence of a single negative eigenvalue of the energy second derivative matrix and the associated negative frequency. The forward ($f$) and reverse ($r$) intrinsic reaction coordinate (IRC) paths were subsequently found where each generated a number of sets of atomic positions correspond to the calculated points on the IRC. The final generated structures on the end of each calculated IRC path was then further geometry-optimized to a local minimum energy structure. All of the IRC calculations were performed with mass-weighted coordinates and the reaction path step-size used in all cases was the default value of 0.1 amu$^{1/2}$-Bohr. Finally, for each point on each IRC path that includes the end minima, single-point calculations were performed to generate the necessary total charge density information. All calculations were performed with the DFT B3LYP/6-31+G(d,p) using Gaussian 09vD.01[38] and were tested for the stability of the generated wave functions to perturbations that included spin restricted-unrestricted perturbations and were found to be stable.

The QTAIM and stress analysis was performed with the AIMAll[39] suite on each wave function obtained in the previous step.

## 4. Results and Discussions

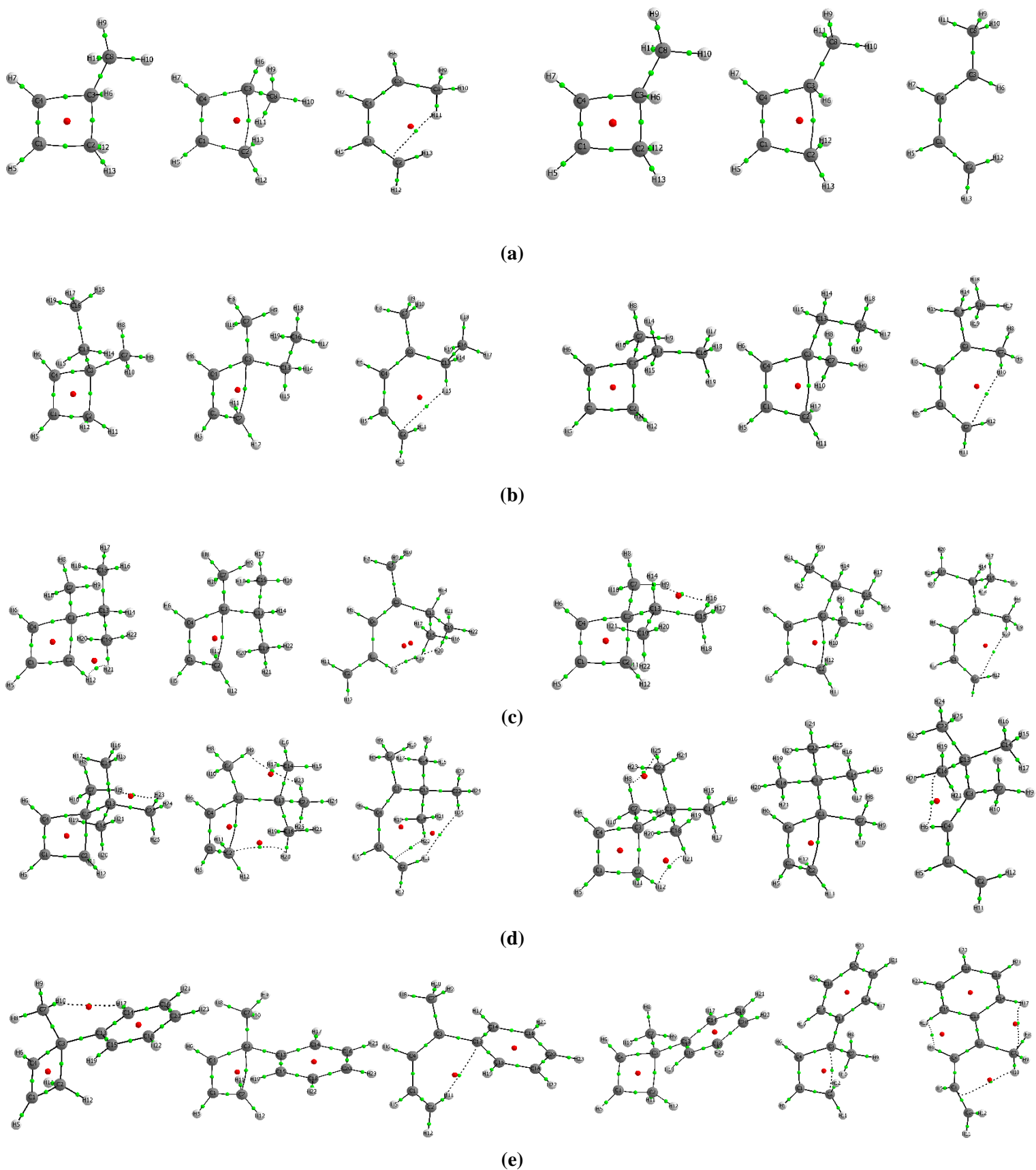

**Figure 1**. Snap-shots of the molecular graphs where the undecorated green and red spheres represented the bond critical points (*BCP*s) and ring critical points (*RCP*s) respectively of the reverse minima, transition state and forward minima are in the left, middle and right panels respectively of each of sub-figures **(a)-(e)**. The ring-opening reactions methyl-cyclobutene (**R1**), ethyl-methyl-cyclobutene (**R2**), Iso-propyl-methyl-cyclobutene (**R3**), ter-butyl-methyl-yclobutene (**R4**) and phenyl-methyl-cyclobutene (**R5**) are presented in sub-figures (**a**)-(**e**) respectively. For each reaction **R1-R5** the inward (TSIC) and outward (TSOC) conrotatory reaction pathways are presented in the left and right-hand (the set of three molecular graphs) panels respectively.

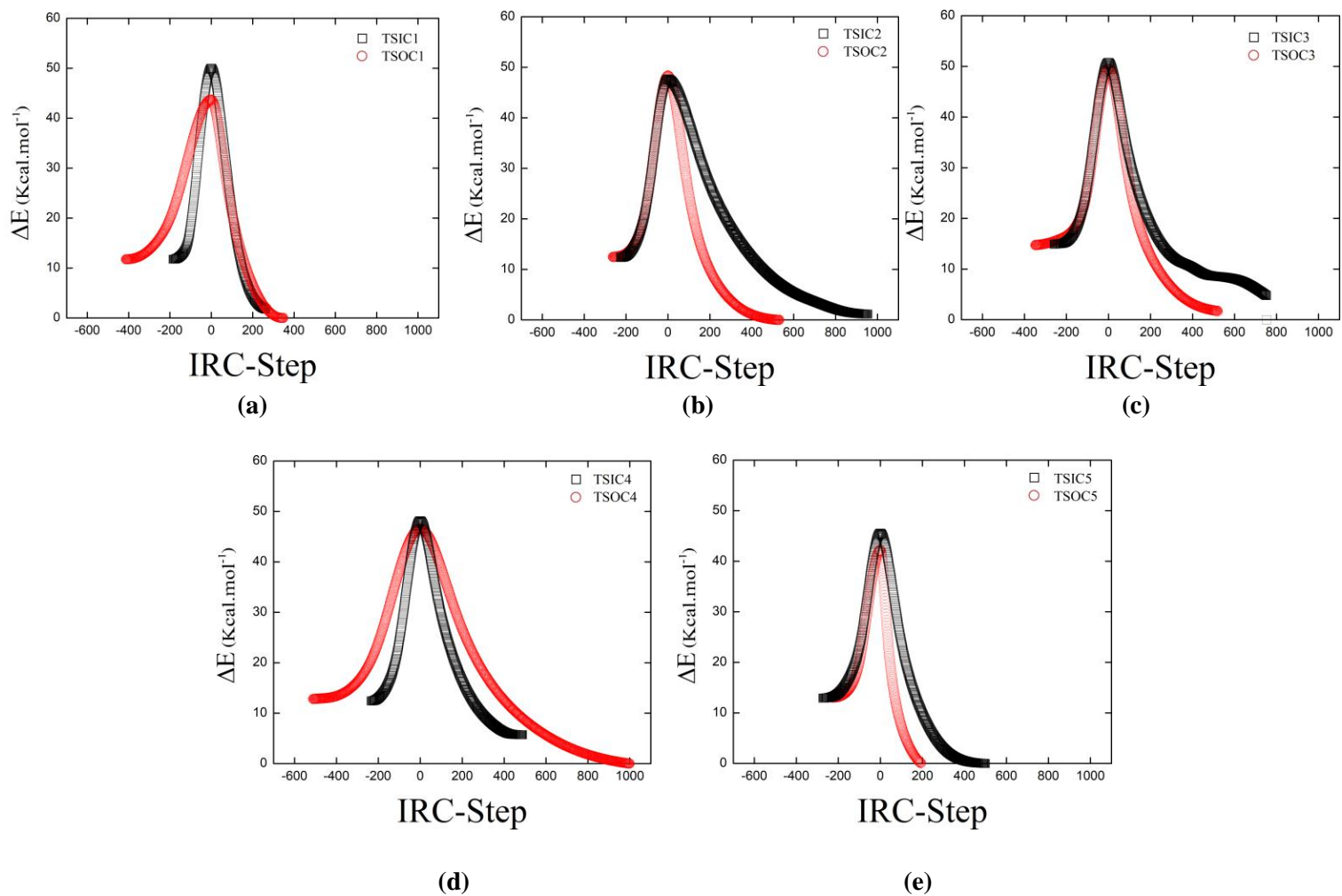

**Figure 2**. The variation of the relative energy ΔE with the IRC are shown for the **R1-R5** reactions in subfigures **(a)**-**(e)** respectively. The larger symbols at the end of each of the IRC represent the energy minima; see the figure caption of **Figure 1** for further details.

*4.1 The character of ring opening reactions from the QTAIM and the stress tensor BCP properties*

In this section we will address the first goal of this investigation; a new criterion to determine whether a reaction is competitive or non-competitive based on chemical properties rather than the somewhat arbitrary ΔTS < 1.0 kcal/mol, see **Table 1** and **Figure 1** and **Figure 2**. Examination of **Table 1** indicates that **R2** is a competitive ring-opening reaction from the criteria of ΔTS < 1.0 kcal/mol whereas the other four reactions; **R1**, **R3** and **R5** are all non-competitive reactions. Visual inspection of the plots of the TSIC and TSOC relative energy ΔE does not immediately indicate which reactions are competitive or non-competitive. In contrast, examination of the plots of the variation of the metallicity $\xi(r_b)$ with the IRC, found for $\xi(r_b) > 1$, for the TSIC and TSOC reactions of the ring opening C2-C3/C2--C3 *BCP* for that **R2**, **R3** and **R5** are degenerate, see **Figure 3(b)** and **Figure 3(c-e)** respectively. This degenerate behavior in the metallicity $\xi(r_b)$ suggests that in fact **R2**, **R3** and **R5** are competitive reactions on the chemical basis of metallicity being key factor to understand the electronic reorganization of the ring opening *BCP* along the reaction pathways. We then see that the corresponding variation of the metallicity $\xi(r_b)$ with the IRC for **R1** and **R4** displays large differences, see **Figure 3(a)** and **Figure 3(d)** respectively. These large differences between the TSIC and TSOC reaction pathways for **R1** and **R4** suggest that **R1** and **R4** are non-competitive reactions. We also notice that the metallicity $\xi(r_b)$, of the ring opening C2-C3/C2--C3 *BCP* for all five reactions **R1-R5**, did not occur at the transition state in confirmation the transition state theory and in agreement with our previous investigation with competitive reactions.[22,40] This clear differentiation between **R2**, **R3** and **R5** as degenerate and **R1**, **R4** is also seen for the other QTAIM measures; ellipticity ε and the total local energy density $H(r_b)$ see the **Supplementary Materials S1** and **Supplementary Materials S2** respectively.

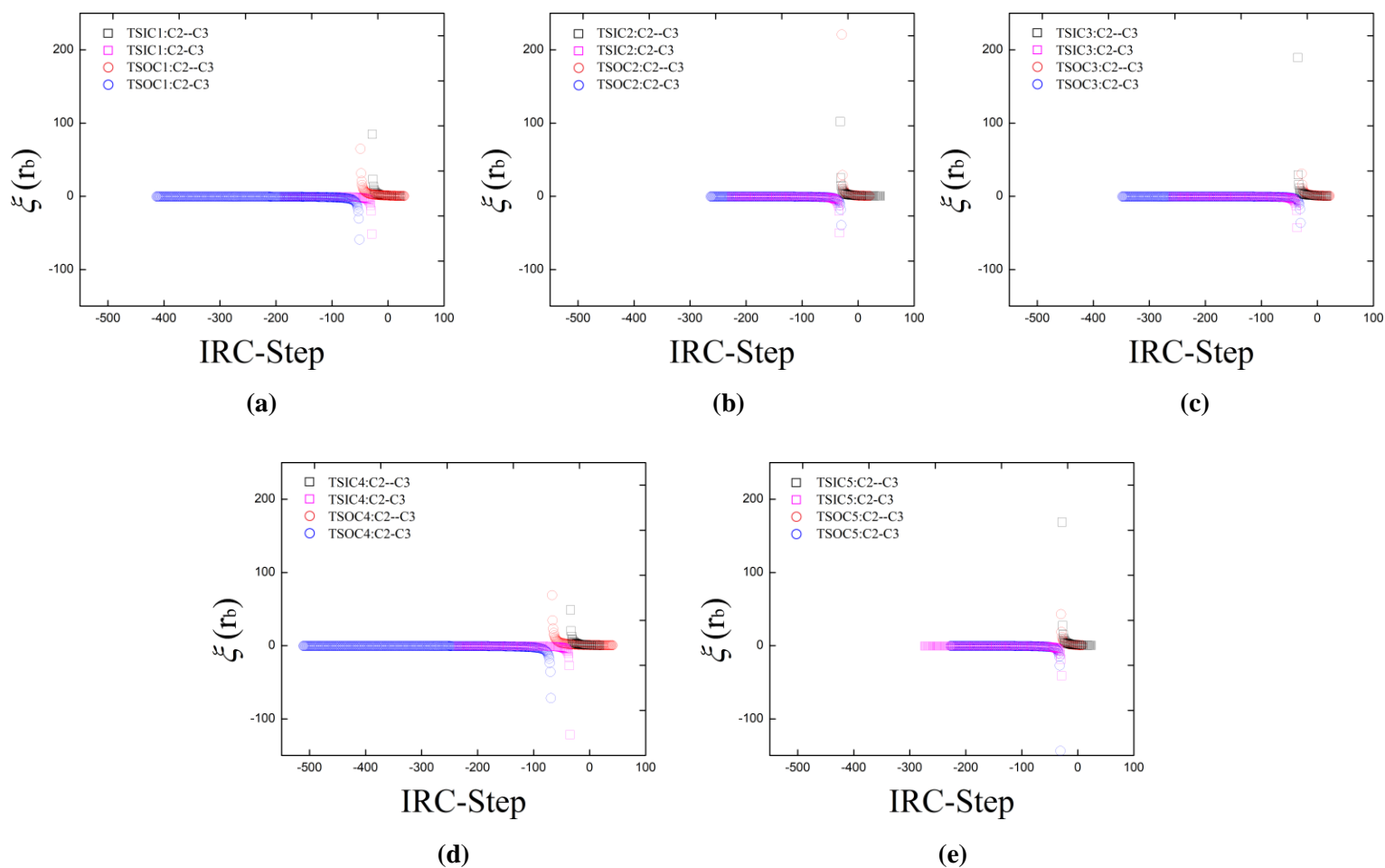

**Figure 3.** The variation of metallicity $\xi(r_b)$ with the IRC of the **R1-R5** reactions for the ring-opening C2-C3 *BCP*/C2--C3 *BCP* is presented in sub-figures **(a)-(e)** respectively, see the figure caption of **Figure 1** for further details. A degree of metallicity is considered present for values of $\xi(\mathbf{r}_b) \geq 1$. The transition of the ring opening C2-C3 *BCP* from a shared-shell *BCP* to a closed-shell C2--C3 *BCP* is indicated the change in sign of the Laplacian $\nabla^2\rho(\mathbf{r}_b) < 0$ to $\nabla^2\rho(\mathbf{r}_b) > 0$.

The stress tensor polarizability $\mathbb{P}_\sigma$ also shows a clear differentiation between **R2**, **R3** and **R5** as degenerate and **R1**, **R4**, see **Figure 4**. This provides another indicator that **R2**, **R3** and **R5** are competitive reactions whereas **R1**, **R4** are non-competitive reactions. In addition, we notice that the transition state does not coincide exactly with the saddle point as was the case for the metallicity $\xi(\mathbf{r}_b)$.

For **R1** and **R4** the position of the maximum in $\mathbb{P}_\sigma$ for the TSOC reaction is located further away from the TS than the maximum in $\mathbb{P}_\sigma$ TSIC reaction; see **Figure 4(a)** and **Figure 4(d)** respectively. We also examined the variation of the stress tensor $\lambda_{3\sigma}$ with the IRC, which exhibits the same trends for **R2**, **R3** and **R5** and **R1**, **R4**, see the **Supplementary Materials S3**. The results for the stress tensor polarizability $\mathbb{P}_\sigma$ and $\lambda_{3\sigma}$ indicate a classification of the **R2**, **R3** and **R5** reactions as competitive reactions and conversely **R1**, **R4** as non-competitive reactions in agreement with the QTAIM results.

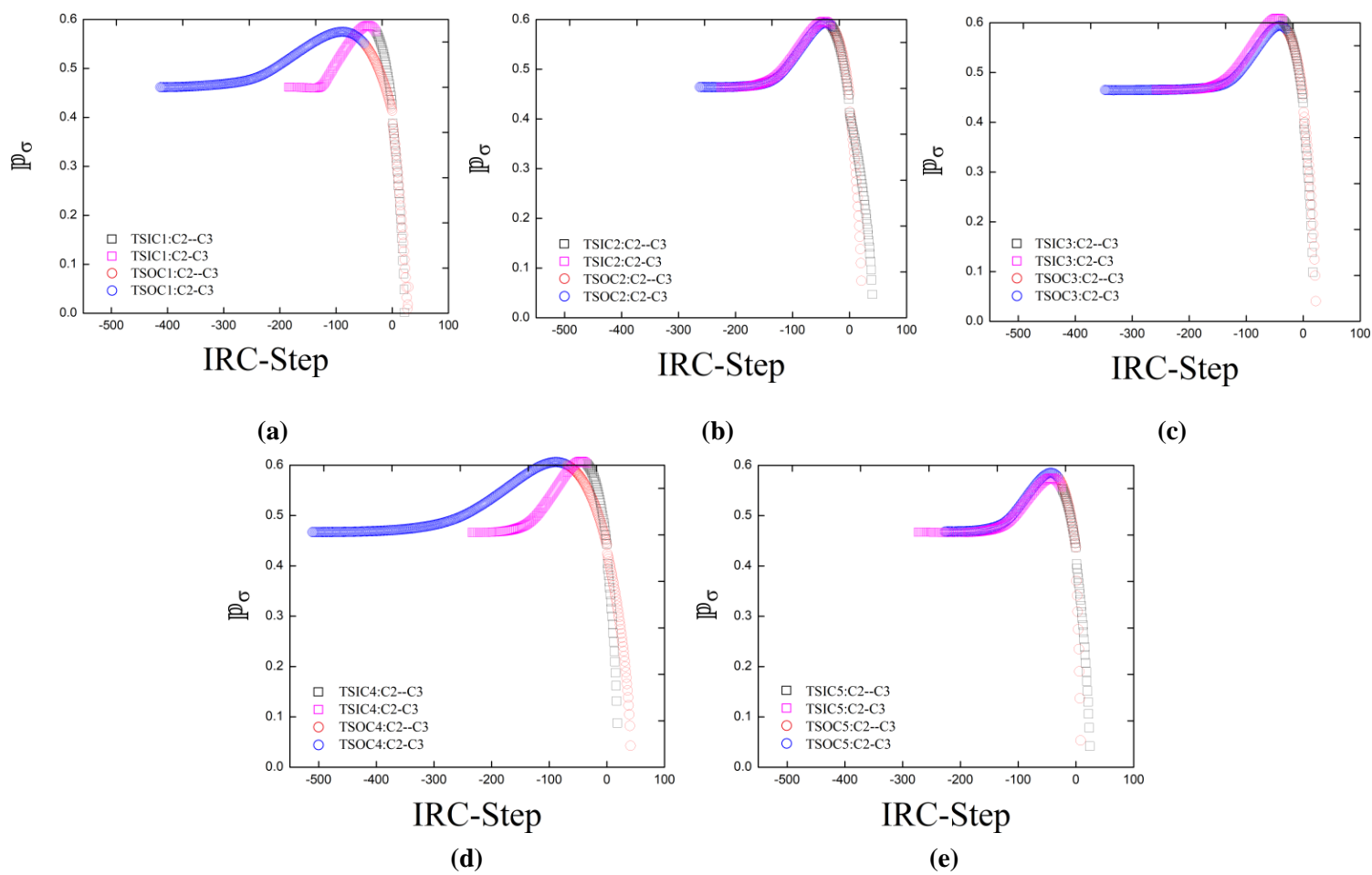

**Figure 4.** The variation of the stress tensor polarizability $\mathbb{P}_\sigma = |\lambda_{3\sigma}|/\lambda_{1\sigma}$ with the IRC of the **R1**-**R5** reactions for the ring-opening C2-C3 *BCP*/C2--C3 *BCP* is included in sub-figures **(a-e)** respectively, see the figure caption of **Figure 1** for further details.

*4.2 The character of ring opening reactions from the QTAIM and the stress tensor bond-path properties*

Examination of the variation of the bond-path length (BPL) with the IRC for the TSIC and TSOC reactions of the ring opening C2-C3/C2--C3 *BCP* shows that **R2**, **R3** and **R5** are degenerate, see **Figure 5(b)** and **Figure 5(c-e)** respectively. In contrast the corresponding variation of the BPL for **R1** and **R4** show large differences, see **Figure 5(a)** and **Figure 5(d)** respectively. Therefore, agreement is found with the QTAIM and stress tensor *BCP* properties that suggest **R2**, **R3** and **R5** reactions as competitive reactions and conversely **R1** and **R4** are non-competitive reactions.

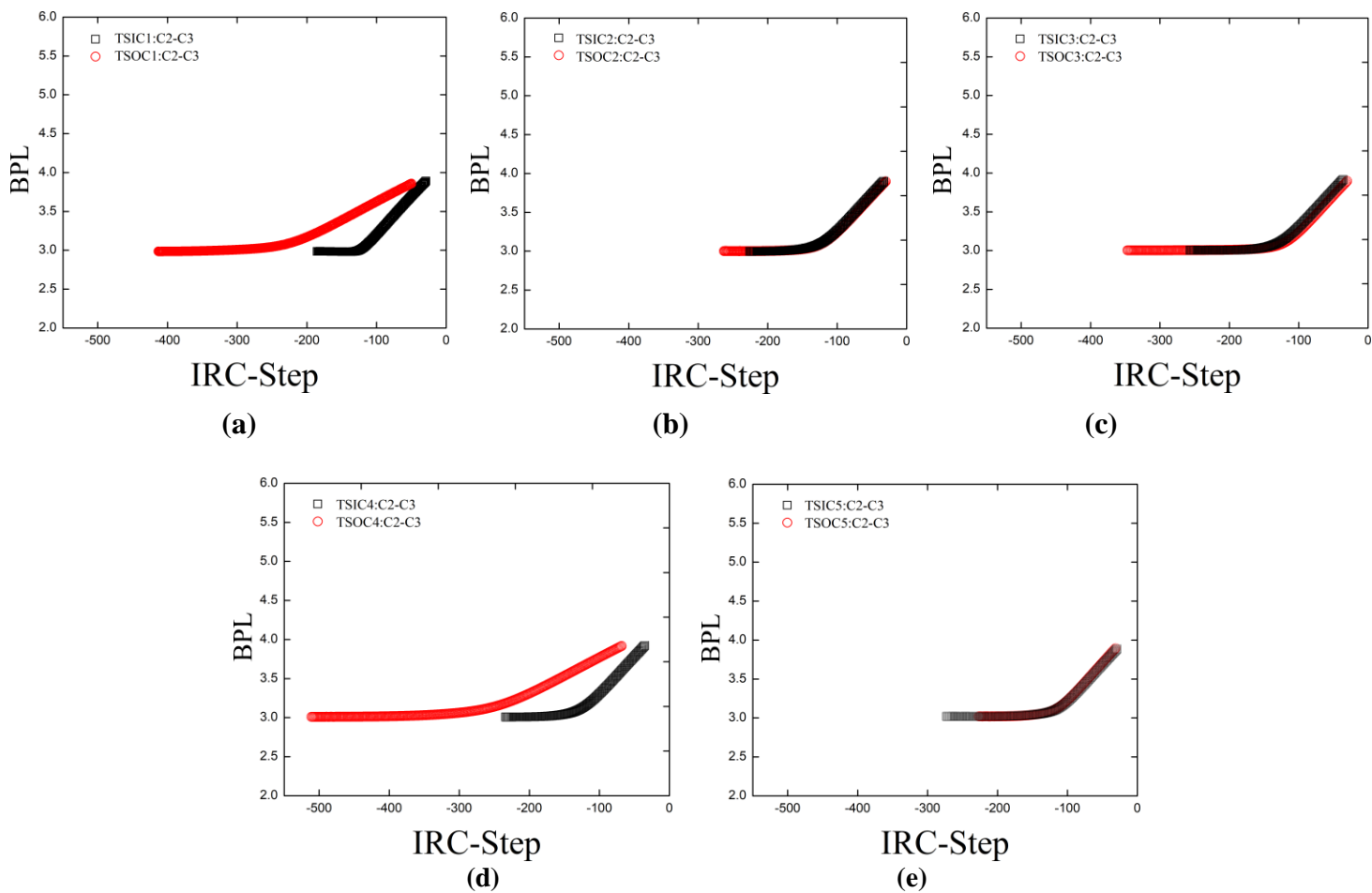

**Figure 5.** The variation of bond-path lengths (in a.u) of the shared-shell ring-opening C2-C3 *BCP* with the IRC of **R1-R5** reactions is presented in sub-figures **(a)-(e)** respectively.

The directional character of the TSIC and TSOC reaction pathways is captured in the bond-path measure the eigenvector-following path length $\mathbb{H}$ since it is constructed from **e**$_2$, the most preferred direction of accumulation of the electronic charge density $\rho(\mathbf{r})$, see the theory section 2.2. Longer eigenvector-following path lengths $\mathbb{H}$ for the shared-shell ring-opening *BCP* of a TSIC vs. TSOC reaction corresponds to a preferred product, TSIC or TSOC. The variation of the $\mathbb{H}$ with the IRC for the TSIC vs. TSOC reactions again shows large differences between **R1** and **R4** that on the basis of the metallicity $\xi(\mathbf{r}_b)$, the stress tensor polarizability $\mathbb{P}_\sigma$ and the BPL that we have previously suggested were non-competitive reactions, see **Figures 3-5** respectively. The use of the eigenvector-following path length $\mathbb{H}$ however, also shows subtle differences between the TSIC and TSOC reaction pathways for **R2**, **R3** and **R5**, see **Figure 6**. Examination of the $\mathbb{H}$ values shows agreement with the experiment yields; **R1**(TSOC), **R2**(TSIC), **R3**(TSIC), **R4**(TSOC) and **R5**(TSOC), see **Figure 6(a-e)** respectively although magnification is required for **R5**, see the inset sub-figure.

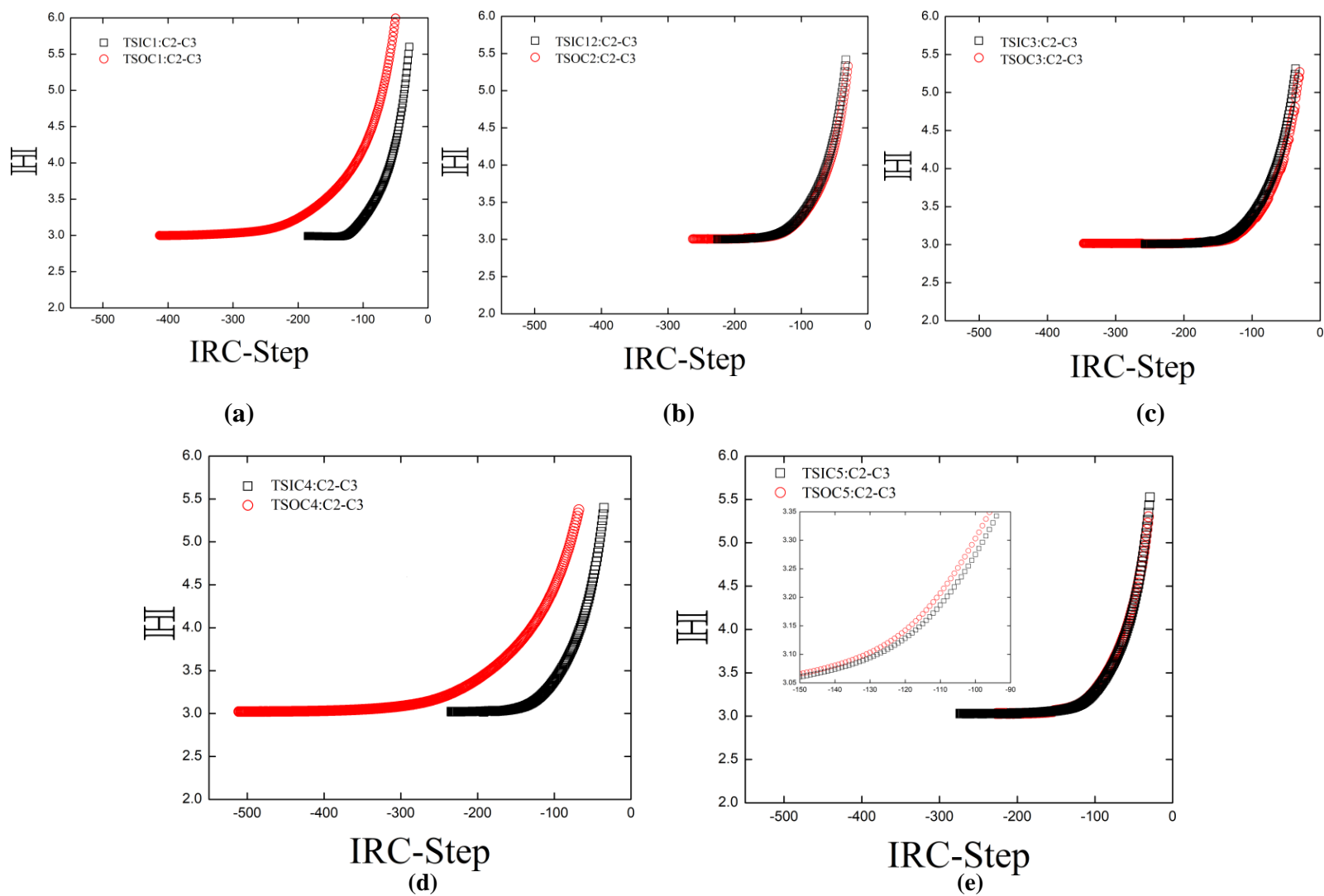

**Figure 6.** The variation of eigenvector-following path length $\mathbb{H}$ (in a.u.) of the shared-shell ring-opening C2-C3 *BCP* with the IRC of **R1-R5** reactions is presented in sub-figures **(a)-(e)** respectively. The corresponding variations of eigenvector-following path length $\mathbb{H}^*$ are provided in the **Supplementary Materials S4**. The procedure for the implementation of the calculation of the eigenvector-following path lengths $\mathbb{H}$ is provided in the **Supplementary Materials S6**.

## 5. Conclusions

In this work, we tested a newly introduced measure, the bond-path framework set $\mathbb{B} = \{p, q, r\}$ from QTAIM as well as using QTAIM and stress tensor measures that we previously developed, to address the two main goals this investigation. The first goal was to find a new chemical, as opposed to energy based criterion to decide if a reaction is competitive or non-competitive. We addressed the first goal by using the scalar QTAIM measures of metallicity $\xi(\mathbf{r_b})$ and the stress tensor polarizability $\mathbb{P}_\sigma$ to determine if a reaction was competitive or non-competitive on the basis of the degeneracy of the results. Overwhelmingly, the **R1** and **R4** reactions were, on the basis of the new chemical as opposed to energy based criterion, determined to be non-competitive reactions and conversely **R2**, **R3** and **R5** were competitive reactions. Agreement was also provided by the total local energy density $H(\mathbf{r_b})$, the stress tensor eigenvalue $\lambda_{3\sigma}$ and the bond-path length (BPL). The use of the calculated activation energies using the somewhat arbitrary measure of $> 1$ kcal/mol does seem to be questionable in the light of the scalar QTAIM and stress tensor results for $\xi(\mathbf{r_b})$ and $\mathbb{P}_\sigma$. These investigations also showed that the maximum values of the metallicity $\xi(\mathbf{r_b})$ and stress tensor polarizability $\mathbb{P}_\sigma$ do not occur at the transition state in agreement with transition state theory and consistent with our earlier investigation of competitive ring-opening reactions.

The second and main goal of this investigation was to find an analysis able to distinguish the torquoselectivity for competitive and non-competitive reaction coordinates that accommodates the directional character of the TSIC and TSOC reaction pathways. This second goal was addressed by using the new QTAIM bond-path framework set $\mathbb{B} = \{p, q, r\}$ where the length of the path $q$ is specified as the eigenvector-following path length $\mathbb{H}$ is constructed from $\underline{\mathbf{e}}_2$, the most preferred direction of accumulation of the electronic charge density $\rho(\mathbf{r})$. In all five reactions examined namely, **R1-R5** the longer lengths $\mathbb{H}$ for the bond-paths of shared-shell ring opening *BCP*s correctly predicted the TSIC or TSOC preferences for the experiment yield. This outcome was more consistent than the use of activation energies (kcal/mol) obtained using b3lyp/6-31G(d,p) theory level, where the calculations failed to correctly predict the correct outcome of the experiment yields for **R3**. The eigenvector-following path length $\mathbb{H}$ values were more successful at distinguishing the torquoselectivity than the scalar QTAIM and stress tensor measures that were only defined at the *BCP*s and lacked a directional component. In addition, the new approach used in this investigation has been demonstrated to be useful for non-competitive as well as competitive reactions and is easier to implement than the previously used directional stress trajectory $\mathbb{T}_\sigma(s)$ analysis that is limited to sets of competitive reactions with very similar nuclear skeletons. The utility of $\mathbb{H}$ is likely because it is constructed from a vector representation of the ellipticity $\varepsilon$ (i.e. path $q$ is a sub-set of the bond-path framework set $\mathbb{B} = \{p, q, r\}$), that contains a necessary directional component, the $\underline{\mathbf{e}}_2$ eigenvector. In addition, the variation of $\mathbb{H}$ for the bond-paths of shared-shell ring opening *BCP*s can follow the subtle changes along the length of entire ring-opening *bond-path* (*r*) as opposed to only following the changes at the *BCP*. More work is needed to test the validity of the $\mathbb{H}$ lengths as predictors of the TSIC or TSOC yield outcomes from experiment. For instance, the preference for TSIC or TSOC is determined from experiment to be a yield of ≈ 70% or more so there are likely limits on the absolute reliability of this new approach in practice.

**Acknowledgements**

The National Natural Science Foundation of China is gratefully acknowledged, project approval number: 21673071. The One Hundred Talents Foundation of Hunan Province and the aid program for the Science and Technology Innovative Research Team in Higher Educational Institutions of Hunan Province are also gratefully acknowledged for the support of S.J. and S.R.K.

**References**

1. M. J. Walker, B. N. Hietbrink, B. E. Thomas, K. Nakamura, E. A. Kallel and K. N. Houk, *J. Org. Chem.*, **2001**, 66, 6669–6672.
2. X.-N. Wang, E. H. Krenske, R. C. Johnston, K. N. Houk and R. P. Hsung, *J. Am. Chem. Soc.*, **2014**, 136, 9802–9805.
3. N. G. Rondan and K. N. Houk, *J. Am. Chem. Soc.*, **1985**, 107, 2099–2111.
4. D. C. Harrowven, M. Mohamed, T. P. Gonçalves, R. J. Whitby, D. Bolien and H. F. Sneddon, *Angew. Chem.*, **2012**, 124, 4481–4484.
5. K. Aikawa, N. Shimizu, K. Honda, Y. Hioki and K. Mikami, *Chem. Sci.*, **2014**, 5, 410–415.
6. Y. Naruse, Y. Ichihashi, T. Shimizu and S. Inagaki, *Org. Lett.*, **2012**, 14, 3728–3731.
7. R. V. López, O. N. Faza and C. S. López, *RSC Adv.*, **2015**, 5, 30405–30408.
8. R. Villar López, O. Nieto Faza and C. Silva López, *RSC Adv.*, **2016**, 6, 59181–59184.
9. B. L. Flynn, N. Manchala and E. H. Krenske, *J. Am. Chem. Soc.*, **2013**, 135, 9156–9163.
10. E. Grenet, J. Martínez and X. J. Salom-Roig, *Chem. - Eur. J.*, **2016**, 22, 16770–16773.
11. L. Akilandeswari and C. Prathipa, *J. Chem. Sci.*, **2015**, 127, 1505–1511.
12. J. E. Barquera-Lozada, *J. Phys. Chem. A*, **2016**, 120, 8450–8460.
13. B. A. Boon, A. G. Green, P. Liu, K. N. Houk and C. A. Merlic, *J. Org. Chem.*, **2017**, 82, 4613–4624.
14. M. Mauksch and S. B. Tsogoeva, *ChemPhysChem*, **2016**, 17, 963–966.
15. H. M. Frey, *Trans. Faraday Soc.*, **1964**, 60, 83–87.
16. M. J. Curry and I. D. R. Stevens, *J. Chem. Soc. Perkin Trans. 2*, **1980**, 0, 1391–1398.
17. A. Morales-Bayuelo, *Int. J. Quantum Chem.*, **2013**, 113, 1534–1543.
18. J. Pilmé, *J. Comput. Chem.*, **2017**, 38, 204–210.
19. A. Morales-Bayuelo, S. Pan, J. Caballero and P. K. Chattaraj, *Phys. Chem. Chem. Phys.*, **2015**, 17, 23104–23111.
20. S. Jenkins, J. R. Maza, T. Xu, D. Jiajun and S. R. Kirk, *Int. J. Quantum Chem.*, **2015**, 115, 1678–1690.
21. S. Jenkins, L. Blancafort, S. R. Kirk and M. J. Bearpark, *Phys. Chem. Chem. Phys.*, **2014**, 16, 7115–7126.
22. H. Guo, A. Morales-Bayuelo, T. Xu, R. Momen, L. Wang, P. Yang, S. R. Kirk and S. Jenkins, *J. Comput. Chem.*, **2016**, 37, 2722–2733.
23. P. S. Lee, X. Zhang and K. N. Houk, *J. Am. Chem. Soc.*, **2003**, 125, 5072–5079.
24. R. F. W. Bader, *Atoms in Molecules: A Quantum Theory*, Oxford University Press, USA, **1994**.

25. R. F. W. Bader, *J. Phys. Chem. A*, **1998**, 102, 7314–7323.
26. R. F. W. Bader, *J. Phys. Chem. A*, **2009**, 113, 10391–10396.
27. E. Kraka and D. Cremer, *J. Mol. Struct. THEOCHEM*, **1992**, 255, 189–206.
28. S. Jenkins, *J. Phys. Condens. Matter*, **2002**, 14, 10251.
29. S. Jenkins, P. W. Ayers, S. R. Kirk, P. Mori-Sánchez and A. Martín Pendás, *Chem. Phys. Lett.*, **2009**, 471, 174–177.
30. P. W. Ayers and S. Jenkins, *Comput. Theor. Chem.*, **2015**, 1053, 112–122.
31. B. Herrera and A. Toro-Labbé, *J. Phys. Chem. A*, **2007**, 111, 5921–5926.
32. R. F. W. Bader, *J. Chem. Phys.*, **1980**, 73, 2871–2883.
33. R. F. W. Bader and T. T. Nguyen-Dang, in *Advances in Quantum Chemistry*, ed. P.-O. Löwdin, Academic Press, **1981**, vol. 14, pp. 63–124.
34. Y. Xu, T. Xu, D. Jiajun, S. R. Kirk and S. Jenkins, *Int. J. Quantum Chem.*, **2016**, 116, 1025–1039.
35. S. Jenkins and M. I. Heggie, *J. Phys. Condens. Matter*, **2000**, 12, 10325.
36. S. Jenkins, *Int. J. Quantum Chem.*, **2013**, 113, 1603–1608.
37. S. Jenkins and I. Morrison, *Chem. Phys. Lett.*, **2000**, 317, 97–102.
38. Michael J. Frisch, G. W. Trucks, H. Bernhard Schlegel, Gustavo E. Scuseria, Michael A. Robb, James R. Cheeseman, Giovanni Scalmani, Vincenzo Barone, Benedetta Mennucci, G. A. Petersson, H. Nakatsuji, M. Caricato, Xiaosong Li, H. P. Hratchian, Artur F. Izmaylov, Julien Bloino, G. Zheng, J. L. Sonnenberg, M. Hada, M. Ehara, K. Toyota, R. Fukuda, J. Hasegawa, M. Ishida, T. Nakajima, Y. Honda, O. Kitao, H. Nakai, T. Vreven, J. A. Montgomery Jr., J. E. Peralta, François Ogliaro, Michael J. Bearpark, Jochen Heyd, E. N. Brothers, K. N. Kudin, V. N. Staroverov, Rika Kobayashi, J. Normand, Krishnan Raghavachari, Alistair P. Rendell, J. C. Burant, S. S. Iyengar, Jacopo Tomasi, M. Cossi, N. Rega, N. J. Millam, M. Klene, J. E. Knox, J. B. Cross, V. Bakken, C. Adamo, J. Jaramillo, R. Gomperts, R. E. Stratmann, O. Yazyev, A. J. Austin, R. Cammi, C. Pomelli, J. W. Ochterski, R. L. Martin, K. Morokuma, V. G. Zakrzewski, G. A. Voth, P. Salvador, J. J. Dannenberg, S. Dapprich, A. D. Daniels, Ö. Farkas, J. B. Foresman, J. V. Ortiz, J. Cioslowski and D. J. Fox, *Gaussian 09, Revision E.01*, Gaussian, Inc., Wallingford, CT, USA, **2009**.
39. T. A. Keith, *AIMAll, Revision 17.01.25*, TK Gristmill Software, Overland Park KS, USA, **2017**.
40. K. J. Laidler, *Chemical kinetics*, Pearson Education Inc, **2009**.

# SUPPLEMENTARY MATERIALS

Predicting Competitive and Non-Competitive Torquoselectivity in Ring-Opening Reactions using QTAIM and the Stress Tensor

Alireza Azizi[1], Roya Momen[1], Alejandro Morales-Bayuelo[2], Tianlv Xu[1], Steven R. Kirk[1*] and Samantha Jenkins[1]*

[1]*Key Laboratory of Chemical Biology and Traditional Chinese Medicine Research and Key Laboratory of Resource Fine-Processing and Advanced Materials of Hunan Province of MOE, College of Chemistry and Chemical Engineering, Hunan Normal University, Changsha Hunan 410081, China.*

[2]*Grupo de Química Cuántica y Teórica de la Universidad de Cartagena, Facultad de Ciencias, Programa de Química, Cartagena de Indias, Colombia.*

e-mail: steven.kirk@cantab.net
e-mail: samanthajsuman@gmail.com

**1. Supplementary Materials S1.** The variation of the ellipticity ε with the IRC of the **R1-R5** reactions for the ring-opening *BCP*.

**2. Supplementary Materials S2.** The variation of the total energy density $H(\mathbf{r_b})$ with the IRC of the **R1-R5** reactions for the ring-opening *BCP*.

**3. Supplementary Materials S3.** The variation of the stress tensor $\lambda_{3\sigma}$ with the IRC of the **R1-R5** reactions for the ring-opening *BCP*.

**4. Supplementary Materials S4.** The variation of eigenvector-following path length $\mathbb{H}^*$ with the IRC of **R1-R5** reactions for the shared-shell ring-opening *BCP*.

**5. Supplementary Materials S5.** The variation of $\mathbb{H}_f$ with the IRC of **R1-R5** reactions for the shared-shell ring-opening *BCP*.

**6. Supplementary Materials S6.** Implementation details of the calculation of the eigenvector-following path lengths $\mathbb{H}$ and $\mathbb{H}^*$.

## 1. Supplementary Materials S1.

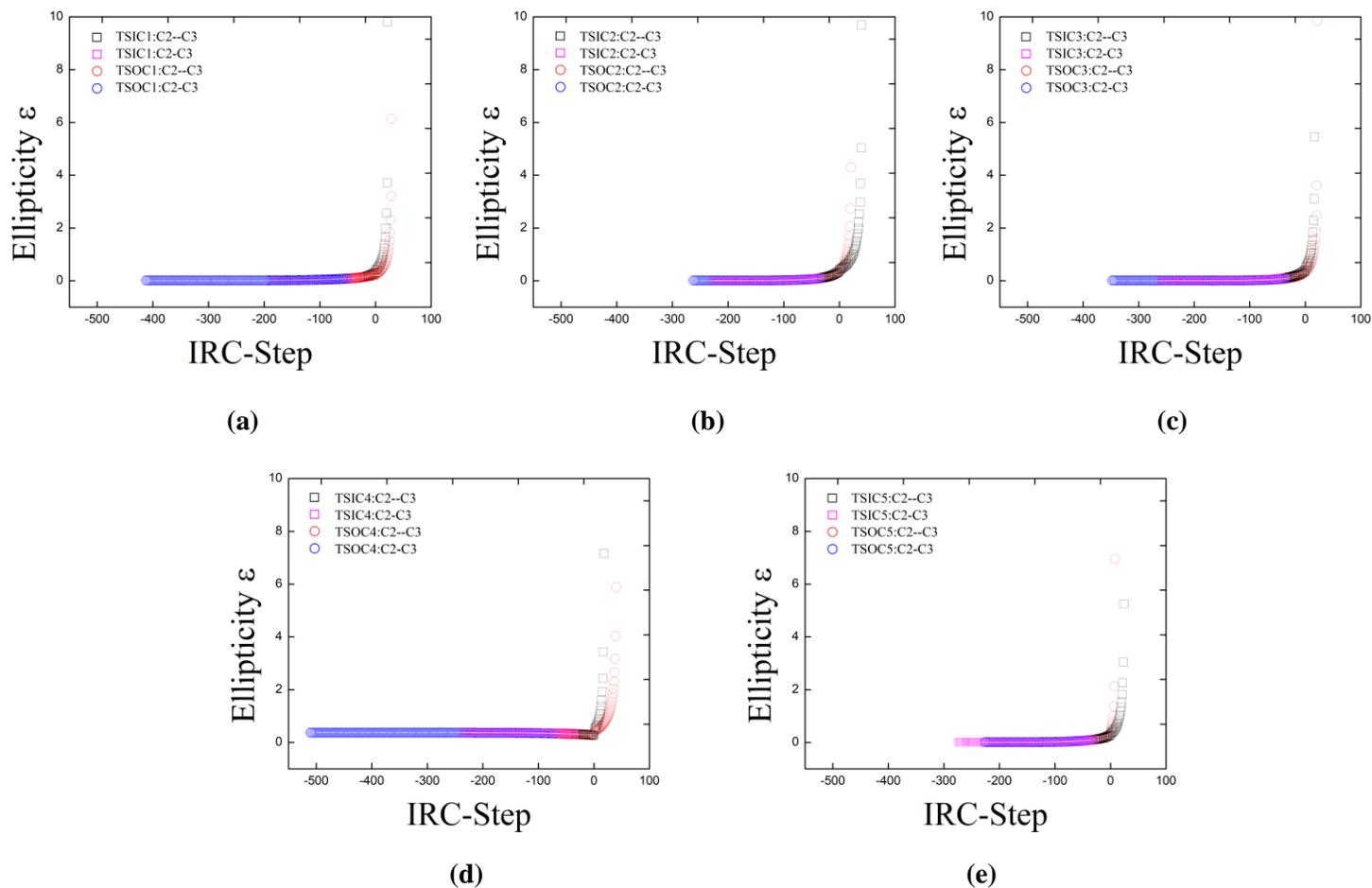

(a)  (b)  (c)

(d)  (e)

**Figure S1.** The variation of the ellipticity ε with the IRC of methyl-cyclobutene (**Reaction-1**), ethyl-methyl-cyclobutene (**Reaction-2**), Iso-propyl-methyl-cyclobutene (**Reaction-3**), ter-butyl-methyl-cyclobutene (**Reaction-4**) and phenyl-methyl-cyclobutene (**Reaction-5**) for the ring-opening C2-C3 *BCP*/C2--C3 *BCP* are presented in sub-figures **(a-e)** respectively.

## 2. Supplementary Materials S2.

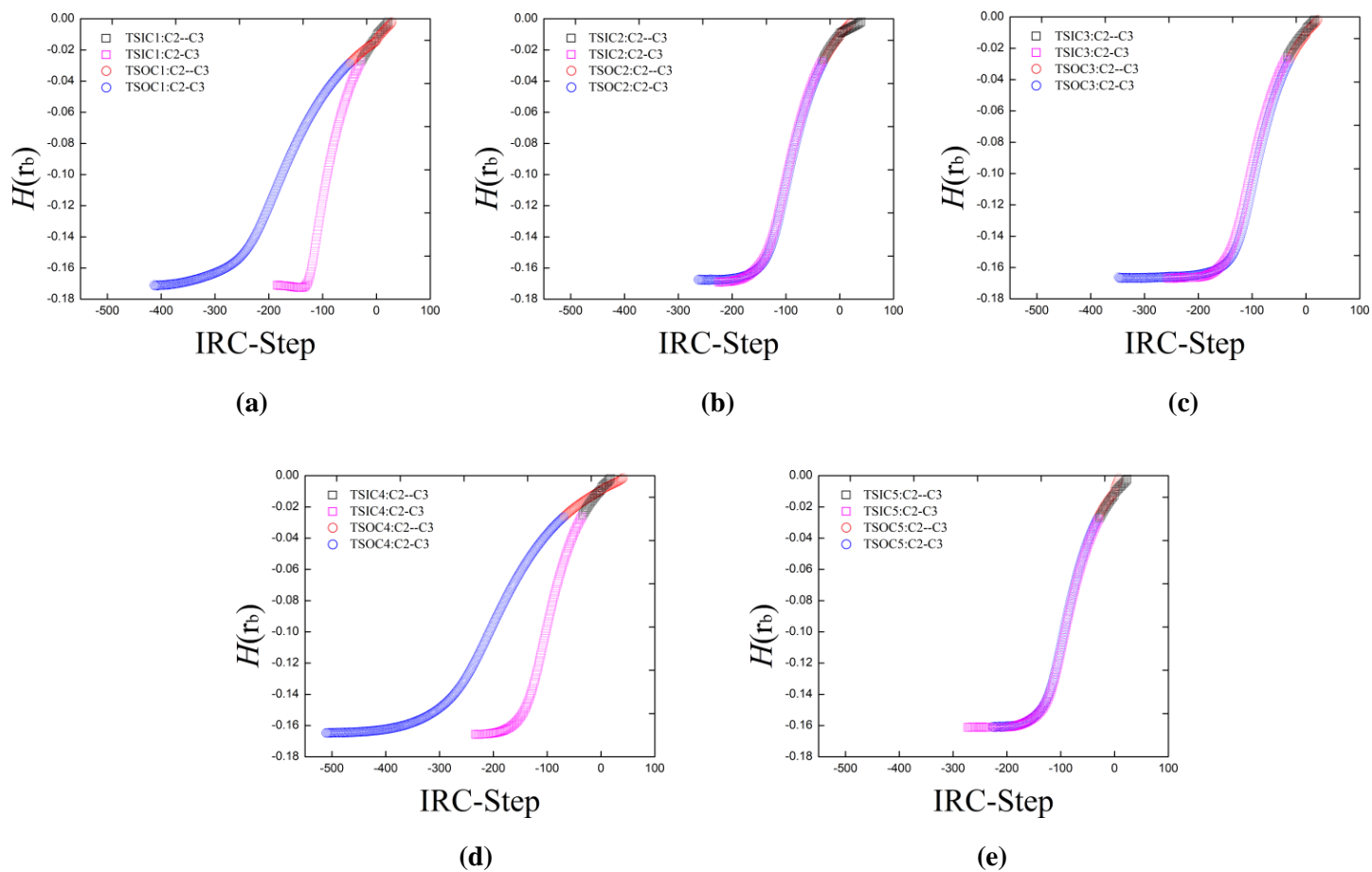

(a) (b) (c)

(d) (e)

**Figure S2.** The variation of the total energy density $H(\mathbf{r_b})$ in (a.u.) with the IRC of the **R1-R5** reactions for the ring-opening C2-C3 *BCP*/C2--C3 *BCP* are presented in sub-figures **(a-e)** respectively. See the figure caption of **Figure S1** for further details.

## 3. Supplementary Materials S3.

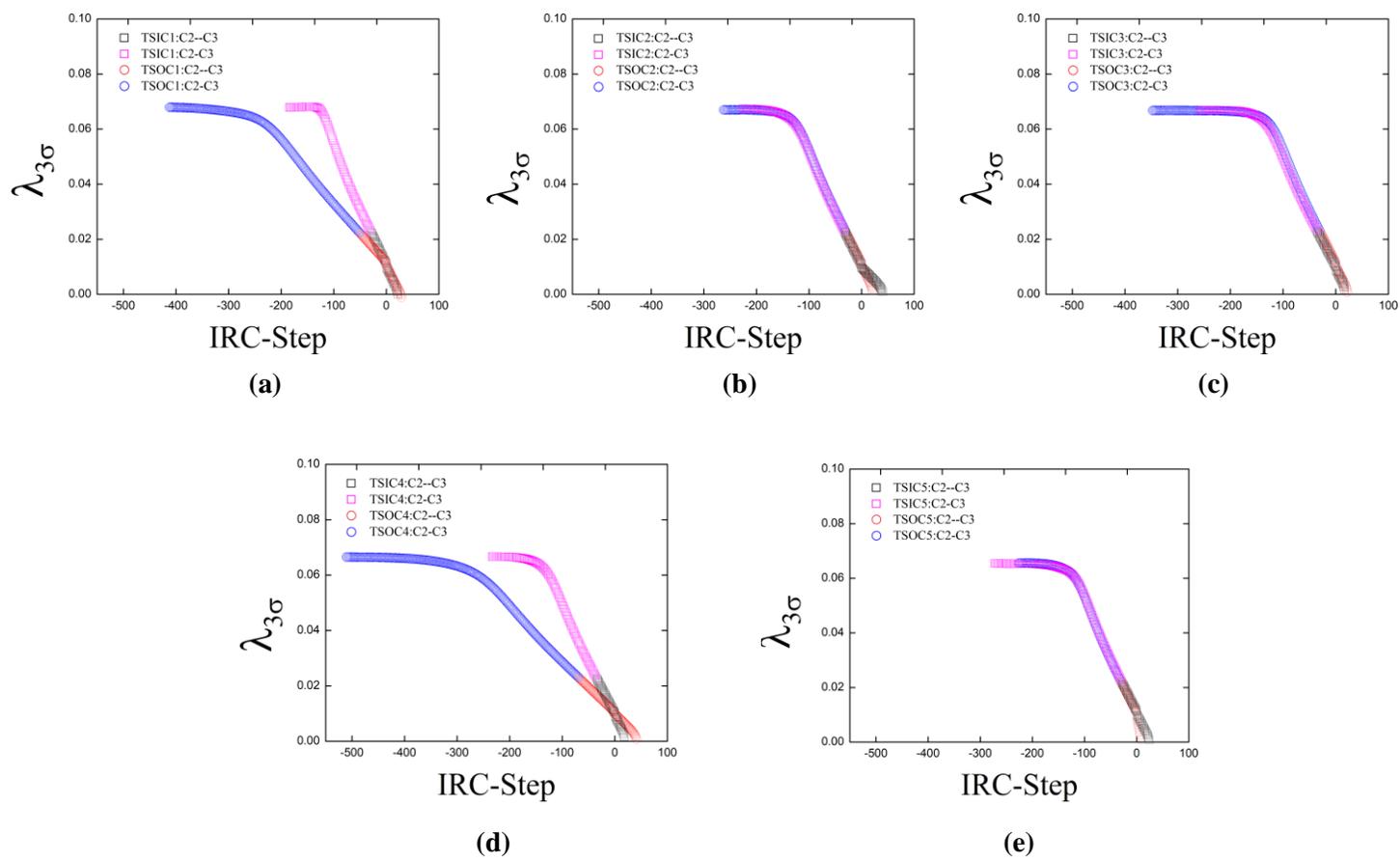

**Figure S3.** The variation of the stress tensor $\lambda_{3\sigma}$ with the IRC of the **R1**-**R5** reactions for the ring-opening C2-C3 *BCP*/C2--C3 *BCP* is included in sub-figures **(a-e)** respectively. See the figure caption of **Figure S1** for further details.

## 4. Supplementary Materials S4.

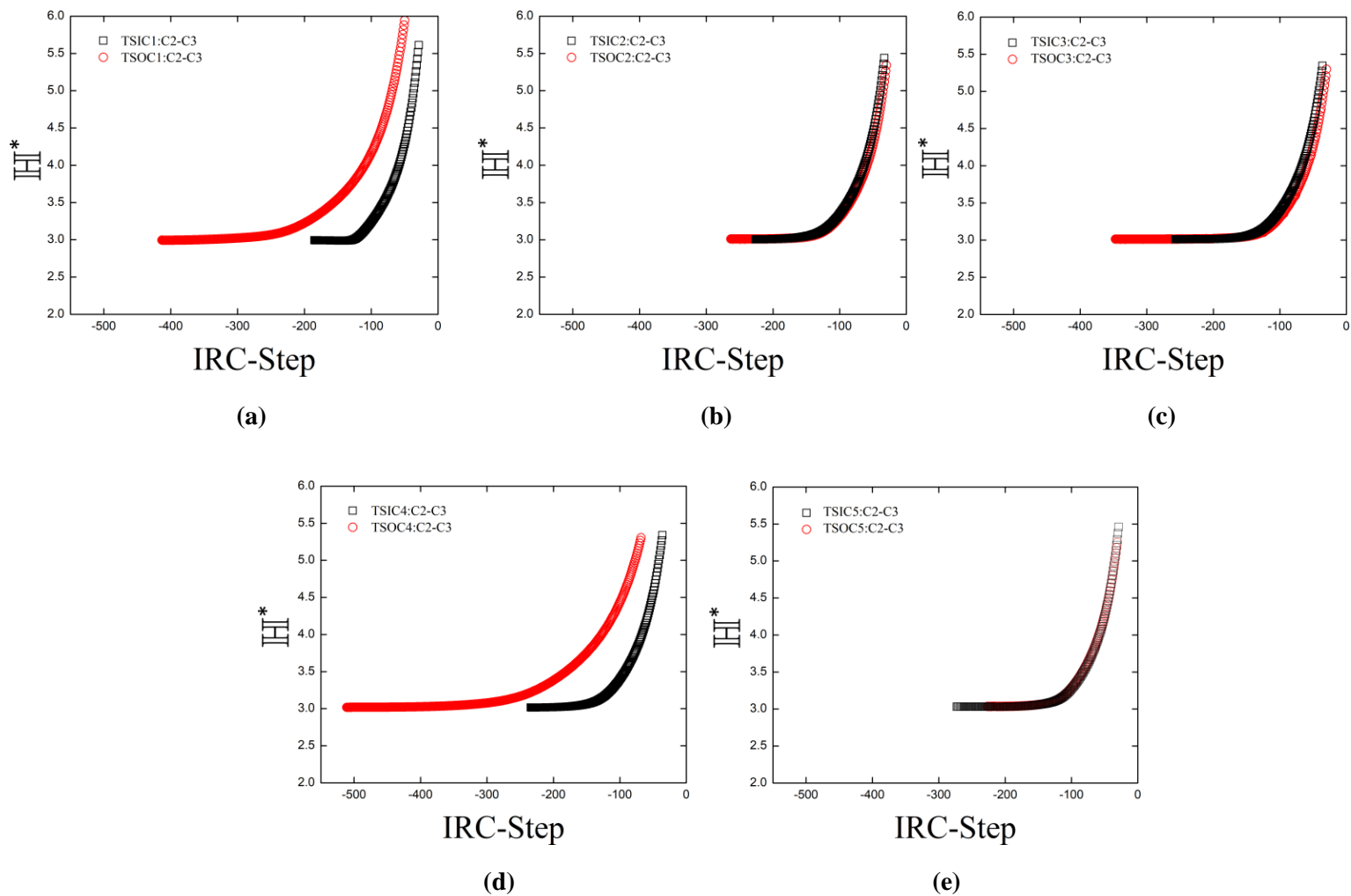

**Figure S4.** The variation of eigenvector-following path length $\mathbb{H}^*$ of the shared-shell ring-opening C2-C3 *BCP* with the IRC of the **R1-R5** reactions is presented in sub-figures **(a)-(e)** respectively. See the figure caption of **Figure S1** for further details.

## 5. Supplementary Materials S5.

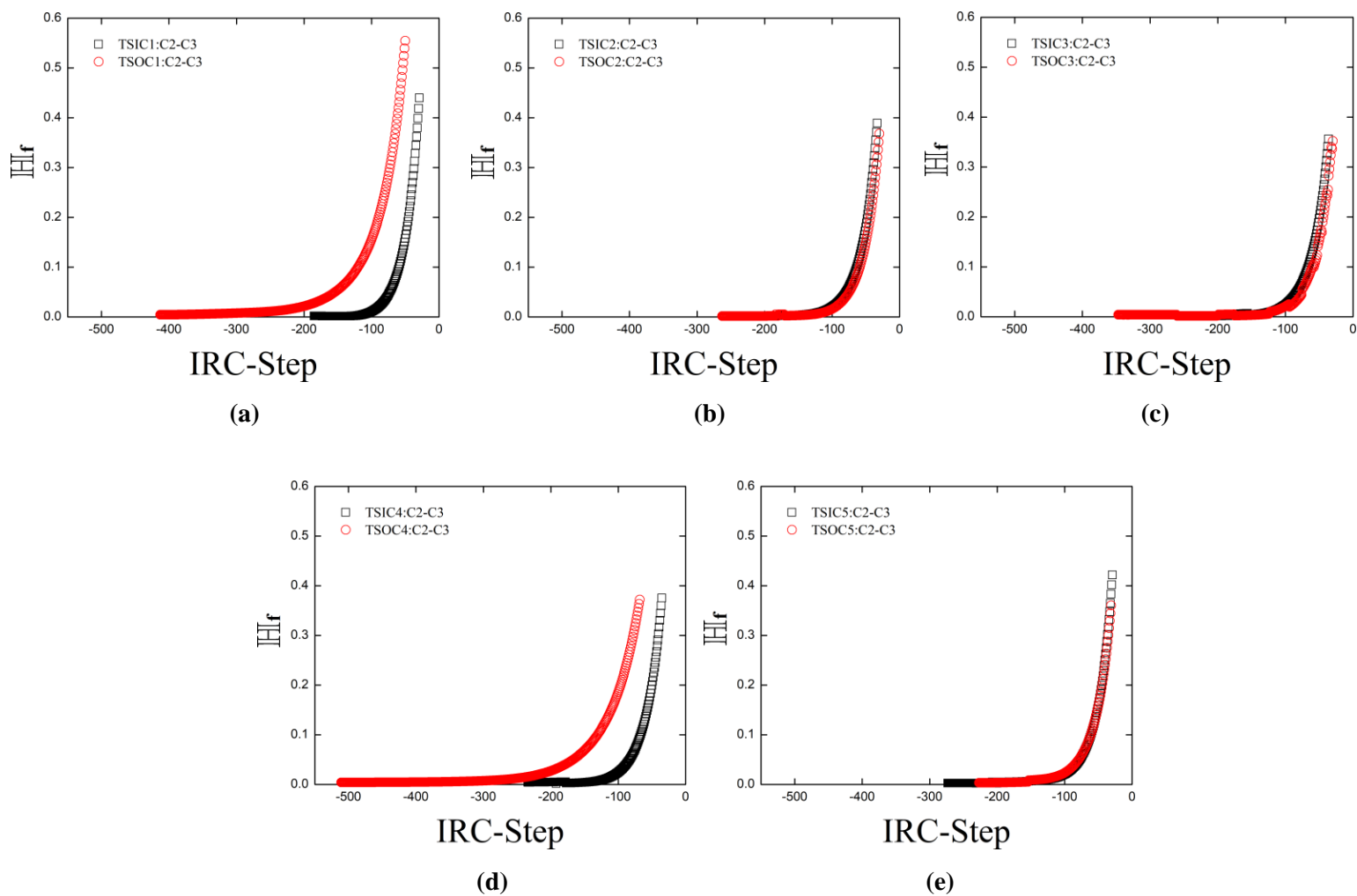

**Figure S5.** The variation of $\mathbb{H}_f$ of the shared-shell ring-opening C2-C3 *BCP* with the IRC of the **R1-R5** reactions is presented in sub-figures **(a)-(e)** respectively. See the figure caption of **Figure S1** for further details.

# 6. Supplementary Materials S6. Implementation details of the calculation of the eigenvector-following path lengths $\mathbb{H}$ and $\mathbb{H}^*$.

When the QTAIM eigenvectors of the Hessian of the charge density $\rho(\mathbf{r})$ are evaluated at points along the bond-path, this is done by requesting them via a spawned process which runs the selected underlying QTAIM code, which then passes the results back to the analysis code. For some datasets, it occurs that, as this evaluation considers one point after another in sequence along the bond-path, the returned calculated $\underline{\mathbf{e}}_2$ (correspondingly $\underline{\mathbf{e}}_1$ is used to obtain $\mathbb{H}^*$) eigenvectors can experience a 180-degree 'flip' at the 'current' bond-path point compared with those evaluated at both the 'previous' and 'next' bond-path points in the sequence. These 'flipped' $\underline{\mathbf{e}}_2$ (or $\underline{\mathbf{e}}_1$) eigenvectors, caused by the underlying details of the numerical implementation in the code that computed them, are perfectly valid, as these are defined to within a scale factor of -1 (i.e. inversion). The analysis code used in this work detects and re-inverts such temporary 'flips' in the $\underline{\mathbf{e}}_2$ (or $\underline{\mathbf{e}}_1$) eigenvectors to maintain consistency with the calculated $\underline{\mathbf{e}}_2$ (or $\underline{\mathbf{e}}_1$) eigenvectors at neighboring bond-path points, in the evaluation of path eigenvector-following path lengths $\mathbb{H}$ and $\mathbb{H}^*$.

.